\documentclass{sigcomm-alternate}
\usepackage{amsfonts,latexsym,amsmath,amstext,amssymb,verbatim,epsfig,psfrag}
\usepackage{colordvi,pstcol}
\usepackage{graphicx,psboxit}
\usepackage{psfrag,graphicx,balance}

\begin{document}
\title{\ttlit D-iteration: evaluation of the update algorithm}

\numberofauthors{1}
\author{
   \alignauthor Dohy Hong\vspace{2mm}\\
   \affaddr{Alcatel-Lucent Bell Labs}\\
   \affaddr{Route de Villejust}\\
   \affaddr{91620 Nozay, France}\\
   \email{dohy.hong@alcatel-lucent.com}
}

\date{\today}
\maketitle

\begin{abstract}
The aim of this paper is to analyse the gain of the update algorithm associated
to the recently proposed D-iteration: the D-iteration is a fluid diffusion based 
new iterative method.
It exploits a simple intuitive decomposition of the product matrix-vector
as elementary operations of fluid diffusion (forward scheme) associated to a new
algebraic representation.
We show through experimentations on real datasets how much this
approach can improve the computation efficiency in presence of the graph
evolution.
\end{abstract}
\category{G.1.3}{Mathematics of Computing}{Numerical Analysis}[Numerical Linear Algebra]
\category{G.2.2}{Discrete Mathematics}{Graph Theory}[Graph algorithms]
\terms{Algorithms, Performance}
\keywords{Computation, Iteration, Fixed point, Gauss-Seidel, Eigenvector.}
\begin{psfrags}
\section{Introduction}\label{sec:intro}
We recall that the PageRank equation can be written under the form:
\begin{eqnarray}\label{eq:affine}
X = P.X + B
\end{eqnarray}
where $P$ is a square non-negative matrix of size $N\times N$ and $B$ a non-negative vector of size $N$.

We recall that the D-iteration approach works when the spectral radius
of $P$ is strictly less than 1 (so that the power series $\sum_{n\ge 0} P^n B$ is convergent).

Based on the ideas proposed in \cite{d-algo, distributed}, we want to
evaluate the efficiency of the D-iteration based update algorithm,
applied on a large matrix (still limited to $N=10^6$ on a single PC) 
in the context of PageRank type equation.

In Section \ref{sec:update}, we recall the general equations and the update
algorithm. Section \ref{sec:PR} presents the web graph dataset based first
analysis.

\section{Update equation}\label{sec:update}
The fluid diffusion model is in the general case described
by the matrix $P$ associated with a weighted graph ($p_{ij}$ is
the weight of the edge from $j$ to $i$) and the initial condition
$F_0$.

We recall the definition of the two vectors used in D-iteration:
the fluid vector $F_n$ by:
\begin{eqnarray}
F_n &=& (I_d - J_{i_n} + P J_{i_n}) F_{n-1}.\label{eq:defF}
\end{eqnarray}
where:
\begin{itemize}
\item $I_d$ is the identity matrix;
\item $I = \{i_1, i_2, ..., i_n,...\}$ with $i_n \in \{1,..,N\}$ is a deterministic or random sequence
  such that the number of occurrence of each value $k\in \{1,..,N\}$ in $I$ is infinity;
\item $J_k$ a matrix with all entries equal to zero except for
  the $k$-th diagonal term: $(J_k)_{kk} = 1$.
\end{itemize}

And the history vector $H_n$ by ($H_0$ initialized to a null vector):
\begin{eqnarray}\label{eq:defH}
H_n &=& \sum_{k=1}^n J_{i_k} F_{k-1}.
\end{eqnarray}

Then, we have (cf. \cite{dohy}):
\begin{eqnarray}\label{eq:H}
H_n + F_n &=& F_0 + P H_n.
\end{eqnarray}

We assume as in \cite{dohy} that the above equation has been computed up to 
the iteration $n_0$ and that
at that time, we are interested to compute the limit associated to $P'$.

Then, it has been shown in \cite{dohy} that the solution of the new limit
$X'$ such that $X' = P' X' + F_0$ can be found by applying 
the D-iteration with $P'$ with modified
initial condition $F_0' = F_{n_0} + (P' - P) H_{n_0}$ and combing its limit
$H_{\infty}'$ to $H_{n_0}$:
$$
H_{n_0} + H_{\infty}' = P' (H_{n_0} + H_{\infty}') + F_0.
$$

Note that the existing iterative methods (such as Jacobi or Gauss-Seidel
iteration, power iteration etc) can naturally adapt the iteration to the modification
of $P$ because they are in general condition initial independent (for any initial vector, the iterative
scheme converges to the unique limit). This is not the of the D-iteration: this is why
while being somehow obvious (once written), the above result is very important.

Such an on-line continuous iteration modification in the context of dynamic
ranking adaptation of PageRank is a priori interesting only
when $P'$ is a minor evolution of $P$. 

In the next section, we analyse the gain of exploiting the information
$H_{n_0}$ and $F_{n_0}$ to compute $H_{\infty}'$ depending on the level of
modification $P'-P$ and a first evaluation of the potential of the distributed
approach with D-iteration described in \cite{distributed}.

\section{Application to PageRank equation: update strategy}\label{sec:PR}
We recall that in the context of PageRank equation, $P$ is of the form:
$dQ$ where $Q$ is a stochastic matrix (or sub-stochastic matrix in presence
of dangling nodes).

For the evaluation purpose, we experimented the D-iteration on a
web graph imported from the dataset \verb+uk-2007-05@1000000+
(available on \verb+http://law.dsi.unimi.it/datasets.php+) which has
41247159 links on 1000000 nodes (0 dangling nodes).

Below we vary $N$ from 1000 to 1000000, extracting from the dataset the
information on the first $N$ nodes.

\begin{table}
\begin{center}
\begin{tabular}{|l|ccc|}
\hline
N & L (nb links) & L/N & D (Nb dangling nodes)\\
\hline
1000 & 12935 & 12.9 & 41 (4.1\%)\\
10000 & 125439 & 12.5 & 80 (0.8\%)\\
100000 & 3141476 & 31.4 & 2729 (2.7\%)\\
1000000 & 41247159 & 41.2 & 45766 (4.6\%)\\
\hline
\end{tabular}\caption{Extracted graph: $N=1000$ to $1000000$.}
\end{center}
\end{table}

\subsection{Modification of link weights}
We call $P(N)$ the initial matrix associated to the $N$ first nodes of the dataset.
Then $P'(N)$ has been built by randomly adding links on the initial graph as follows:
\begin{itemize}
\item Scenario S(m): each node $i$ is selected with probability  $mL\epsilon/N$
  then $m$ link is added to a random destination node $j$.
\end{itemize}
When $\epsilon$ is very small, it adds in average $m\epsilon\times 100$\% links on the existing
$L$ links and it can not create more than $m\times N$ links.
Note that with such a process we can create two links between node $i$ and $j$: 
it should be understood then as a modification of the link weight.
Note also that when there were $n$ outgoing links from $i$, an addition of a new link
from $i$ modifies the $n$ previous ones (from $1/n$ to $1/(n+1)$).

\subsection{Analysis of the evaluation}
\subsubsection{N=1000}
We first consider $N=1000$ case with S(1).
\begin{figure}[htbp]
\centering
\includegraphics[angle=-90, width=\linewidth]{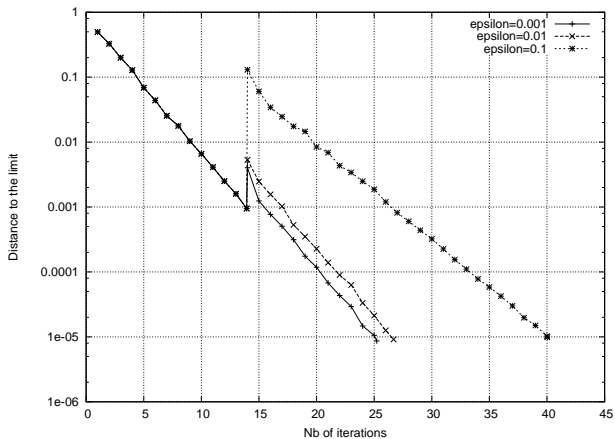}
\caption{Scenario S(1) with $N=1000$: adding 12, 115 and 1000 links.}
\label{fig:S1-1000}
\end{figure}

Figure \ref{fig:S1-1000} shows the evolution of the distance to the limit:
$P$ is applied up to the time we get a distance to the limit of $1/N = 0.001$,
then $P'$ is introduced following the scenario S(1). The iteration number (x-axis)
is defined as the number of links that was used (a diffusion from one node to
$n$ nodes is counted as $n$ links utilization) divided by $L$, so that 1 iteration
is almost equivalent to the cost of the product of $P$ by a vector (roughly, one iteration
of the power iteration).
It says that the perturbation of
\begin{itemize}
\item $0.1$\% (12 links added) requires about 1.5 iterations to find back the precision
  before the perturbation, which means roughly 90\% of previous computation are reused (about 10 times faster
  than restarting);
\item $1$\% (115 links created) requires about 3 iterations to find back the precision
  before the perturbation: it clearly shows that the impact of the link creation depends
  on its location and on the graph structure;
\item $10$\% (1000 links created, which is the maximum) requires about 12 iterations
 to find back the precision of $1/N$, which means roughly 30\% of previous computation are reused.
\end{itemize}
At this stage, let's just keep in mind that the impact of $x$\% modification of links
may be sensitive.

\begin{figure}[htbp]
\centering
\includegraphics[angle=-90, width=\linewidth]{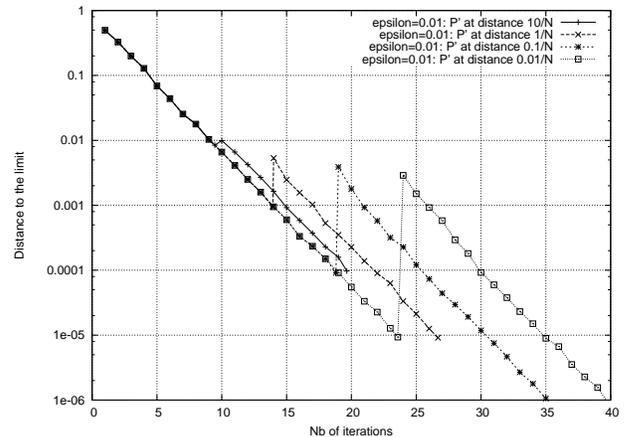}
\caption{Scenario S(1) with $N=1000$: varying the introduction moment of $P'$.}
\label{fig:S1-1000-0.01}
\end{figure}

It is in fact easy to understand that the distance between $X$ and $X'$ depends
on the degree of modification: Figure \ref{fig:S1-1000-0.01} shows that when the
115 links ($1$\% case) are added, $|X-X'|$ ($L_1$ norm) is about 0.005, which makes
useless all computations with precision below 0.005 with $P$.

\begin{figure}[htbp]
\centering
\includegraphics[angle=-90, width=\linewidth]{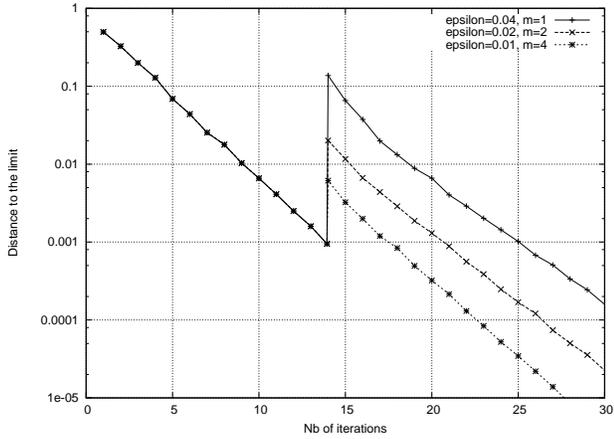}
\caption{$N=1000$: impact of $m$.}
\label{fig:S-1000-0.04-var}
\end{figure}
Figure \ref{fig:S-1000-0.04-var} shows the comparison of the distance $|X-X'|$
when about 500 links are created by: ($\epsilon=0.04$ and $m=1$) or ($\epsilon=0.02$ and $m=2$)
or ($\epsilon=0.01$ and $m=4$). In respective cases, we added 508, 502 and 484 links.

Clearly the impact is higher when modifying more different nodes.

\subsubsection{N=10000}
We restart here the same analysis with the 3 figures with
$N=10000$.
Figure \ref{fig:S1-10000} shows here somehow similar results than with $N=1000$, but globally with
a better reuse of previous computations:
here, the perturbation of
\begin{itemize}
\item $0.1$\% (129 links added) requires about 3 iterations
 to find back the precision of $1/N$;
\item $1$\% (1250 links created) requires about 4 iterations
 to find back the precision of $1/N$;
\item $10$\% (10000 links created, which is the maximum) implies that about 50\% of previous computation
 effort is reused for the computation of the new limit.
\end{itemize}

\begin{figure}[htbp]
\centering
\includegraphics[angle=-90, width=\linewidth]{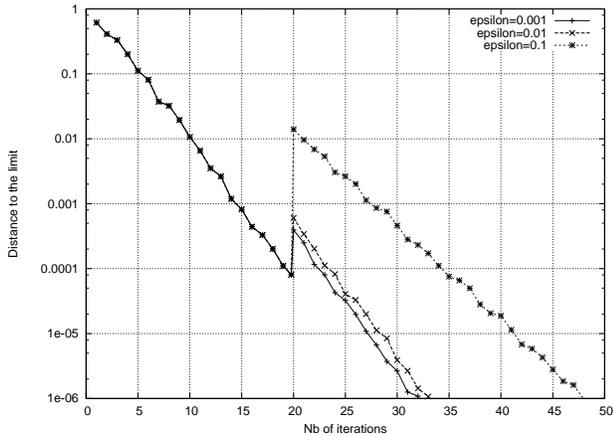}
\caption{Scenario S(1) with $N=10000$: adding 129, 1250 and 10000 links.}
\label{fig:S1-10000}
\end{figure}

The results of Figure \ref{fig:S1-10000-0.01} is as expected.
Here, the distance $|X-X'|$ is about 0.001.
\begin{figure}[htbp]
\centering
\includegraphics[angle=-90, width=\linewidth]{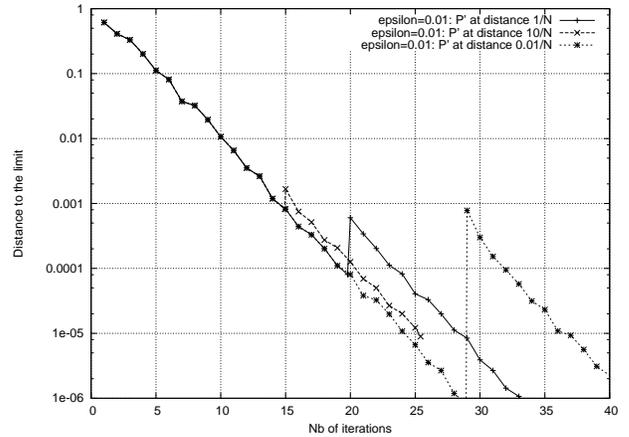}
\caption{Scenario S(1) with $N=10000$: varying the introduction moment of $P'$.}
\label{fig:S1-10000-0.01}
\end{figure}

A possible first explanation of a more or less important impact of the link modification
on $|X-X'|$ is the ratio of the dangling nodes: indeed, when a link is
added to a dangling node, it will transfer to its child most of its score, bringing
more differences for $X'$. This partially explains the better reuse of the previous computations
for $X$ with $N=10000$.

\subsubsection{N=100000}

\begin{figure}[htbp]
\centering
\includegraphics[angle=-90, width=\linewidth]{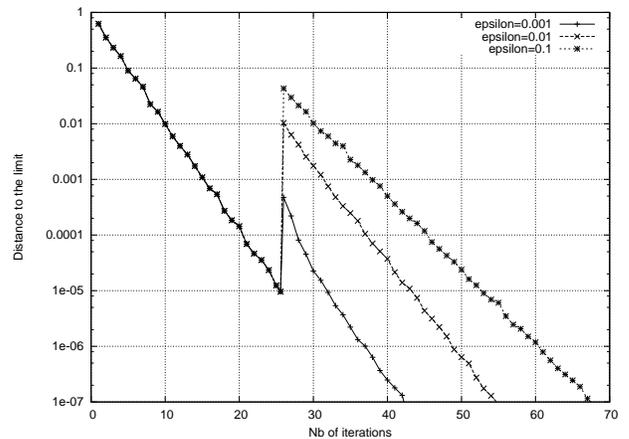}
\caption{Scenario S(1) with $N=100000$: adding 3037, 31593, 100000 links.}
\label{fig:S1-100000}
\end{figure}
With $N=100000$, we observe in Figure \ref{fig:S1-100000} a jump size
that's more expected: we roughly expect the ratios of 10 and 3 between successive peaks at iteration 26
corresponding to the ratios of the number of links added.

\subsubsection{N=1000000}
\begin{figure}[htbp]
\centering
\includegraphics[angle=-90, width=\linewidth]{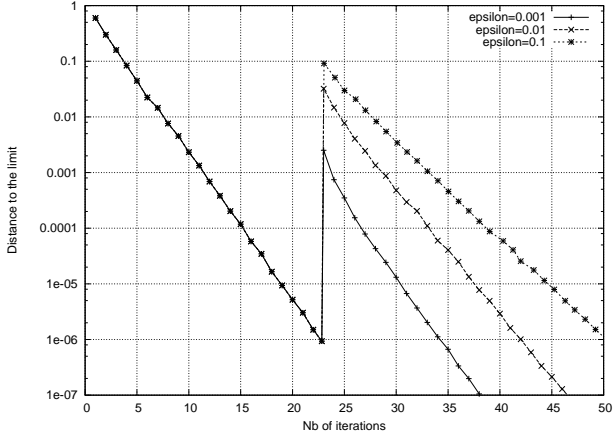}
\caption{Scenario S(1) with $N=1000000$: adding 41259, 412191, 1000000 links.}
\label{fig:S1-1000000}
\end{figure}
With $N=1000000$, we observe somehow unexpected results: Figure \ref{fig:S1-1000000} 
shows that the addition of $0.1$\% links may create a huge update cost: merely less than 
50\% of previous computations are reused for $X'$.

\subsection{Adding new nodes}
The effect of adding new nodes is a bit different from the above modification, because
it brings a global modification of the initial fluid $F_0$.
We considered here a very simple operation of nodes addition without any links: as
a consequence, we have to modify only the current value of $F_n$ adding on the existing
nodes: $(1-d)(1/(N+1) - 1/N)$ and on the new node a fluid of $1/(N+1)$.

\subsubsection{N=1000}

The effect of nodes addition is shown in Figure \ref{fig:S1-1000-node}: we can
see that their effect is quite comparable to the addition of $x$\% links.
\begin{figure}[htbp]
\centering
\includegraphics[angle=-90, width=\linewidth]{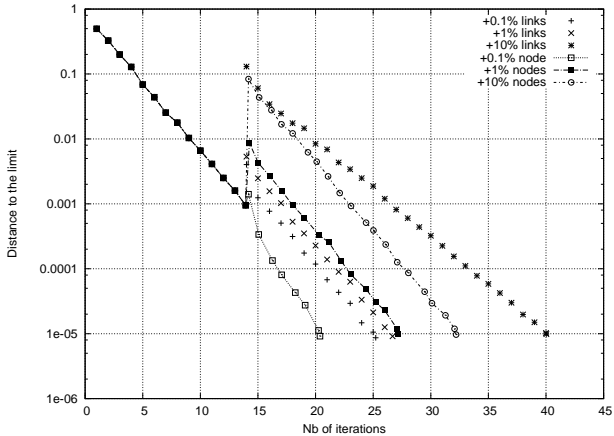}
\caption{Adding 1 (0.1\%), 10 (1\%) and 100 (10\%) new nodes to $N=1000$ nodes.}
\label{fig:S1-1000-node}
\end{figure}

\subsubsection{N=10000}
\begin{figure}[htbp]
\centering
\includegraphics[angle=-90, width=\linewidth]{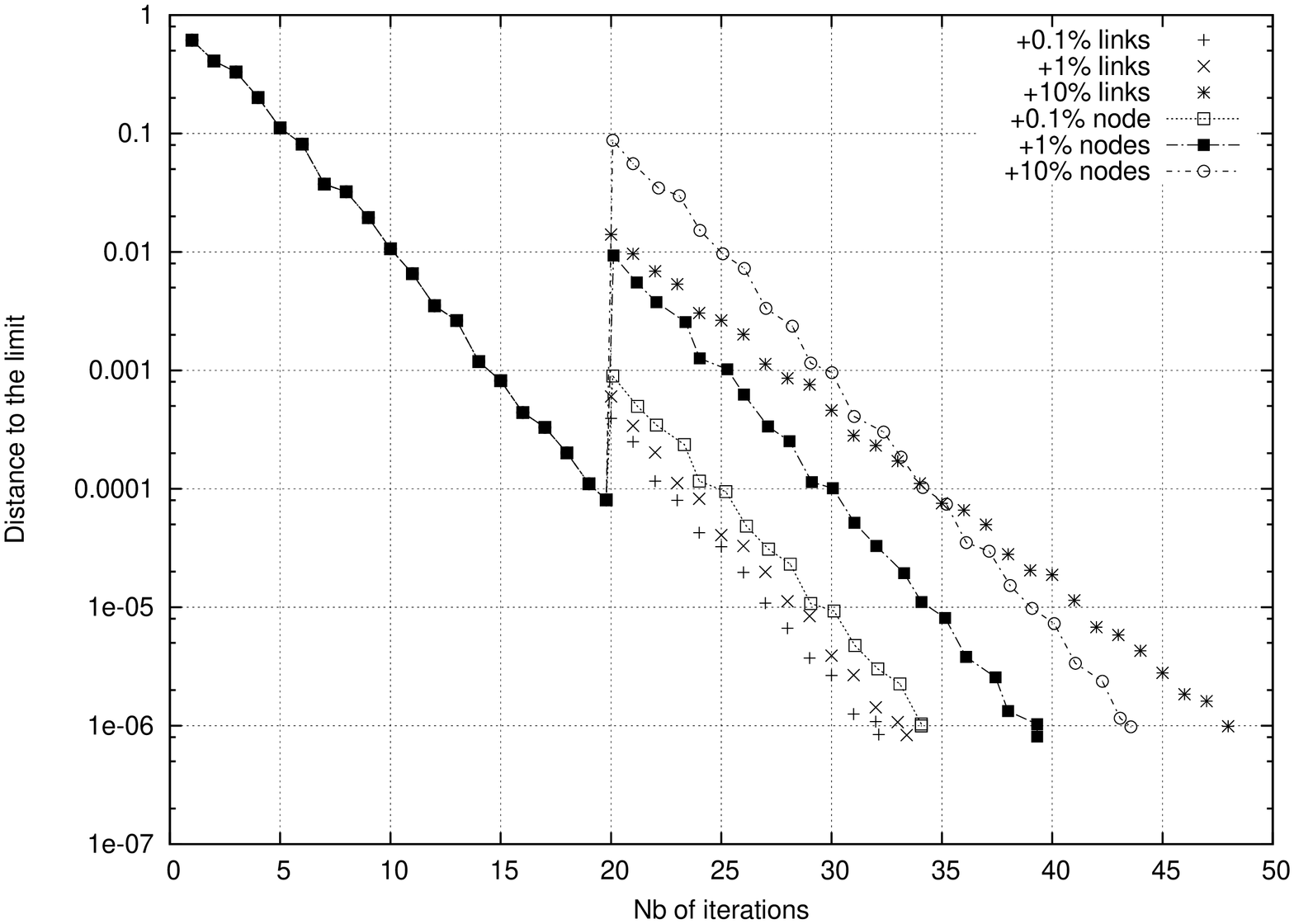}
\caption{Adding 10 (0.1\%), 100 (1\%) and 1000 (10\%) new nodes to $N=10000$ nodes.}
\label{fig:S1-10000-node}
\end{figure}

\subsubsection{N=100000}
\begin{figure}[htbp]
\centering
\includegraphics[angle=-90, width=\linewidth]{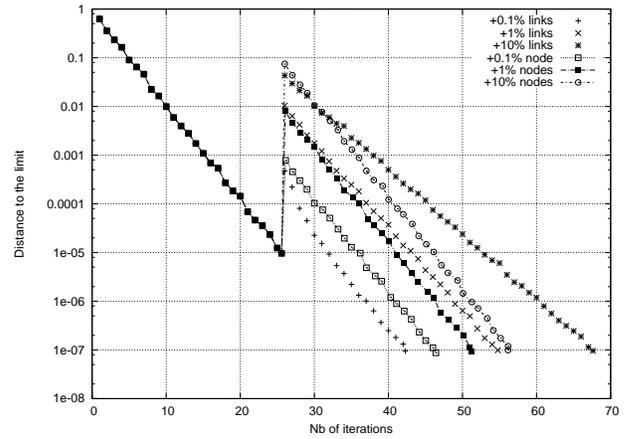}
\caption{Adding 100 (0.1\%), 1000 (1\%) and 10000 (10\%) new nodes to $N=100000$ nodes.}
\label{fig:S1-100000-node}
\end{figure}

The above results are much easier to understand then links addition:
when $x$\% of nodes are added, this roughly creates a total of $x/100$ fluid, so that
whatever the value of the initial $N$, the jumps for $P'$ are at distance 0.1, 0.01 and 0.001
(resp. for 0.1\%, 1\% and 10\% nodes addition).
Then, the jumps for the links modification can be understood as a variation of nodes addition
plus a modification of the convergence slope (the slope modification is of course the consequence
of the matrix $P$ modification). 

\subsection{Conclusion of the analysis}
A first quick conclusion of the above results is that the impact of the link modification
on the eigenvector limit may be much more than the proportion of the modification:
this is expected and there are probably here two reasons. One is the fact that
the PageRank can be inherited to the children iteratively (obvious reason) and second,
the because of the PageRank formulation, adding a link on a node with $n$ outgoing links,
the total weight modification is doubled: $2/(n+1)$. 
However, we saw that, as a first approximation, 
a modification of $x$\% of links leads to an impact
corresponding to an addition of $x$\% nodes.  
As a consequence, when the size of $N$ becomes larger, the gain of the update strategy
is automatically reduced.

\section{Conclusion}\label{sec:conclusion}
In this paper, we showed through simple analysis that the D-iteration method based
update strategy has a limited gain, 
when $N$ becomes larger, compared to the full computation of the eigenvector
when the transition matrix $P$ evolves in time.

\end{psfrags}
\bibliographystyle{abbrv}
\bibliography{sigproc}

\end{document}